\documentclass[aps,prc,groupedaddress,showpacs,manuscript]{revtex4}

\usepackage{graphicx}

\begin{document}

\title{Medium modifications of the nucleon-nucleon elastic cross section in neutron-rich intermediate energy HICs}

\author {Qingfeng Li$\, ^{1}$\footnote{Fellow of the Alexander von Humboldt Foundation.}
\email[]{Qi.Li@fias.uni-frankfurt.de}, Zhuxia Li$\, ^{2}$
\email[]{lizwux@iris.ciae.ac.cn}, Sven Soff$\, ^{3}$, Marcus
Bleicher$\, ^{3}$, and Horst St\"{o}cker$\, ^{1,3}$}
\address{
1) Frankfurt Institute for Advanced Studies (FIAS), Johann Wolfgang Goethe-Universit\"{a}t, Max-von-Laue-Str.\ 1, D-60438 Frankfurt am Main, Germany\\
2) China Institute of Atomic Energy, P.O.\ Box 275 (18),
Beijing 102413, P.R.\ China\\
3) Institut f\"{u}r Theoretische Physik, Johann Wolfgang Goethe-Universit\"{a}t, Max-von-Laue-Str.\ 1, D-60438 Frankfurt am Main, Germany\\
 }


\begin{abstract}
Several observables of unbound nucleons which are to some extent
sensitive to the medium modifications of nucleon-nucleon elastic
cross sections in neutron-rich intermediate energy heavy ion
collisions are investigated. The splitting effect of neutron and
proton effective masses on cross sections is discussed. It is
found that the transverse flow as a function of rapidity, the
$Q_{zz}$ as a function of momentum, and the ratio of halfwidths of
the transverse to that of longitudinal rapidity distribution
$R_{t/l}$ are very sensitive to the medium modifications of the
cross sections. The transverse momentum distribution of
correlation functions of two-nucleons does not yield information
on the in-medium cross section.
\end{abstract}


 \maketitle

\section{Introduction}
The isospin dependence of the in-medium nucleon-nucleon (NN)
interaction in a dense neutron-rich nuclear matter attracts more
and more interest with the development of upcoming experiments at
the Rare Isotope Accelerator (RIA) laboratory (USA) and at the new
international accelerator Facility for Antiproton and Ion Research
(FAIR) at the Gesellschaft f\"{u}r Schwerionenforschung (GSI,
Germany). Within transport theory both of the mean field and the
in-medium two-body scattering cross sections eventually come from
the same origin and can derived from the same NN interaction
\cite{Han94,Bass98}. In \cite{Han94} the in-medium NN elastic
scattering cross sections were studied based on the quantum
hadrodynamics (QHD) model and the Skyrme interaction with
closed-time Green's function technique without considering the
isospin dependence of cross sections. The symmetry potential
energy in the mean field part has been explored and found to be
very important for the understanding of many problems in
intermediate energy nuclear physics as well as in astrophysics
(see, for example, \cite{BaoAnBook01,baranRP,Li:1997px}).
Naturally, the next step is to ask about the isospin dependence of
the in-medium NN cross section and how to probe it practically.

The isospin dependence of the in-medium NN elastic cross sections
was studied based on the extended QHD model in which $\rho$
\cite{Li:2000sh} as well as $\delta$ \cite{Li:2003vd} mesons are
included. In \cite{Li:2003vd} it was found that $\sigma_{nn}^*$ is
smaller than that of $\sigma_{pp}^*$ since the effective neutron
mass in neutron-rich matter is smaller than that of proton after
considering the contribution of $\delta$ meson \cite{Liu:2001iz}.
Recently, the neutron and proton mass splitting are widely
studied. There exist two typically different definitions on the
effective mass: the Dirac mass $m_D^*$ and the nonrelativistic
effective mass (NR) $m_{NR}^*$
\cite{DiToro:2005ac,vanDalen:2005ns}. They actually have complete
different origin. And, it is further found that the neutron's
Dirac mass is always smaller than the proton's in a neutron-rich
medium. For the NR mass, the situation becomes more complicated,
Dalen et al \cite{vanDalen:2005ns} showed that within the
relativistic mean field theory (RMF) the NR mass has the same
behavior with Dirac mass but when the NR mass is calculated with
Dirac-Brueckner-Hartree-Fock (DBHF) theory \cite{vanDalen:2005ns,
Sammarruca:2005ch,Sammarruca:2005tk} and nonrelativistic
Brueckner-Hartree-Fock (BHF) theory \cite{Zuo:2001bd,Zuo:2005hw}
the non-relativistic neutron mass becomes larger than that of
protons in the neutron-rich medium. When the NR mass is calculated
with Skyrme interactions at mean field level, it is found that
whether the neutron mass larger or smaller than protons depends on
the version of Skyrme interaction used, for instance, the neutron
effective mass is larger than proton's for $SKM^{*}$ while for
SLyb, it is just opposite \cite{DiToro:2005ac}.

Recently the sensitive probes to the medium modification of NN
elastic cross sections in neutron-rich heavy ion collisions (HICs)
at intermediate energies was studied by B.A. Li et al
\cite{Li:2005ib,Li:2005jy}, in which the NR mass splitting was
used. Again, it is seen that the effective cross sections are
influenced contrarily by the different definitions of effective
nuclear mass: based on the effective Lagrangian of density
dependent relativistic hadron theory, our calculations
\cite{Li:2003vd} give the trend (in neutron-rich nuclear medium):
$\sigma_{nn}^*< \sigma_{pp}^*$; while, based on the DBHF model
\cite{Sammarruca:2005ch,Sammarruca:2005tk}, the trend is on the
contrary: $\sigma_{nn}^*> \sigma_{pp}^*$. For the effective $np$
elastic cross section $\sigma_{np}^*$, two approaches give the
similar result. As a default in this work, we address the
splitting $\sigma_{nn}^*<\sigma_{pp}^*$ as the "Dirac" case while
the $\sigma_{nn}^*>\sigma_{pp}^*$ as the "NR" case (supposing a
neutron-rich medium). How does the difference between the
in-medium NN cross sections resulting from different splitting
effects (NR and Dirac cases) influences the dynamics of HICs at
intermediate energies? What (more) observables are sensitive to
the medium modification of NN elastic cross sections? In this
paper we would like to extend this topic with more observables.

The new updated UrQMD transport model especially for simulating
the intermediate energy HICs \cite{Li:2005zz,Li:2005kq,Li:2005gf}
is adopted for calculations in this work. A soft equation of state
(EoS) with a symmetry potential energy depending linearly on the
nuclear density is adopted in this work \cite{Li:2005gf}. The
neutron-rich reactions $^{96}{\rm Zr}+^{96}{\rm Zr}$ and
$^{78}{\rm Ni}+^{96}{\rm Zr}$ at a beam energy $E_b=100A\,{\rm
MeV}$ and for reduced impact parameters $b/b_0=0$ and $0.5$ are
chosen, where $b_0=R_{proj}+R_{targ}$ is the maximum impact
parameter for the colliding system. For each reaction $2\cdot
10^5$ events are calculated.

In this work, we suppose that the in-medium cross sections can be
factorized as the product of a medium correction factor
($F(u,\alpha,p)$) and the free NN elastic scattering cross
sections ($\sigma^{free}$) based on present results of theoretical
calculations in Refs.
\cite{{Sammarruca:2005tk},{Sammarruca:2005ch},{Li:2005jy},Li:2003vd},
which reads
\begin{equation}
\sigma^*=F(u,\alpha,p) \sigma^{\rm free}. \label{ftot}
\end{equation}
The medium correction factor $F$ depends on the nuclear reduced
density u (=$\rho/\rho_0$), the isospin-asymmetry
$\alpha=(\rho_n-\rho_p)/\rho$ and momentum, generally. As usual,
the $\rho$, $\rho_n$, and $\rho_p$ are the nuclear, neutron and
proton densities, and the $\rho_0$ represents the normal nuclear
density, respectively. In order to study various effects we
consider
 three different cases here, i.e., (1), for
"NoMed", $F=1$, which means that we use the cross sections in free
space; (2), for "PartMed", $F=F_\alpha\cdot F_u$, which means that
we consider the isospin-scalar density effect ($F_u$) and the
isospin-vector splitting effect ($F_\alpha$) on the NN elastic
cross sections; and (3), for "FullMed", $F=F_\alpha^{\rm p}\cdot
F_u^{\rm p}$, which means that we further consider the momentum
constraints on the case "PartMed", the density dependence of the
splitting effect is also considered. The $F_u$ and $F_\alpha$ are
expressed as follows

\begin{equation}
F_u=\frac{1}{3}+\frac{2}{3}\exp[-u/0.54568], \label{fr}
\end{equation}
which is similar to the density dependence of the scaling factor
used in \cite{Li:2005jy}. From Eq. (\ref{fr}) the decrease of
cross sections as a function of nuclear density is clear, for
example, $F_{u=2}=0.35$. It was also seen from our previous work \cite{Li:2003vd} that the density dependence of neutron-neutron (or proton-proton) and neutron-proton
elastic cross sections is also different when we consider the isospin vector $\rho$-meson contribution. This effect is not considered here in order to observe more clearly the probable splitting effect which is originally from the isospin vector $\delta$-meson contribution from the point of view of the extended QHD theory.

To model the splitting effect in in-medium nucleon-nucleon elastic
cross sections, we use

\begin{equation}
F_\alpha=1+\tau_{ij}\eta A(u)\alpha. \label{fd}
\end{equation}
For $\tau_{ij}$ in Eq. (\ref{fd}), when $i=j=n$, $\tau_{ij}=-1$;
$i=j=p$, $\tau_{ij}=+1$; and when $i \neq j$, $\tau_{ij}=0$. The
$\eta = +1$ and $-1$ represent the Dirac and NR typed splittings,
respectively. $A(u)$ represents the density dependence of the
splitting effect $F_\alpha$, which is different between Dirac and
NR cases and expressed as

\begin{equation}
A(u)=\left\{
\begin{array}{l}
0.85 \hspace{1.5cm}  {\rm "PartMed-NR"}  \\
\frac{0.85}{1+3.25 u} \hspace{1cm} {\rm "FullMed-NR"}
\end{array}
\right. ,\label{fau1}
\end{equation}
and
\begin{equation}
A(u)=\left\{
\begin{array}{l}
0 \hspace{1.5cm}  {\rm "PartMed-Dirac"}  \\
0.25 u \hspace{1cm} {\rm "FullMed-Dirac"}
\end{array}
\right. ,\label{fau2}
\end{equation}
respectively. The different density dependence of $A(u)$ shown in
Eqs. (\ref{fau1}) and (\ref{fau2}) originates from the different
density dependence of the splitting effect on cross sections based
on different theories \cite{Li:2003vd,Sammarruca:2005ch}, i.e., In
\cite{Sammarruca:2005ch} it was seen that the sensitivity of the
splitting effect of neutron-neutron and proton-proton elastic
cross sections to the isospin asymmetry is weaker at larger
densities, while in our previous work based on the extended QHD
model \cite{Li:2003vd} an increasing density dependence of the
splitting effect was seen due to a larger neutron and proton
effective mass-splitting at higher densities. The different
behavior of the density dependence of the splitting effect on
cross sections is interesting and deserves further investigation.

The $F_\alpha^{\rm p}$ and $F_u^{\rm p}$ factors in the case
"FullMed" are expressed in one formula,
\begin{equation}
F_{\alpha,u}^{\rm p}=\left\{
\begin{array}{l}
1 \hspace{3.3cm} p_{NN}>1 {\rm GeV}/c \\
\frac{F_{\alpha,u} -1}{1+(p_{NN}/0.425)^5}+1 \hspace{1cm} p_{NN}<1 {\rm GeV}/c

\end{array}
\right. , \label{fdpup}
\end{equation}
with $p_{NN}$ being the relative momentum in the NN center-of-mass
system. It is seen that the density- and splitting- effects on the
NN elastic cross section (in Eqs. (\ref{fr})-(\ref{fau2})) will
disappear at high momenta such as $p_{NN}>1\, {\rm GeV}/c$  which
was implied in \cite{Sammarruca:2005ch,Li:2005jy} for the NR case
. For comparison, this study adopts the same momentum constraint
for the Dirac splitting case as for the NR case.

In addition, the temperature effect on the nucleon-nucleon cross
sections should be considered. Unfortunately, the theoretical
predictions of this effect are not very robust. For example, by
using a thermodynamic Green's function approach with
nonrelativistic propagators and with the ladder approximation for
the thermodynamic $T$ matrix, a sharp resonance structure of the
in-medium cross section is present at low temperatures.
Furthermore, with a smaller nuclear density, this cusplike
behavior is more distinct \cite{Alm:1994db}. Nevertheless, based
on the extend QHD (QHD II) model and introducing temperature
dependent distribution functions of fermions and anti-fermions, we
found that the effective nucleon-nucleon elastic cross section
increases slowly with the increase of the temperature
\cite{Li:2003vd}. Therefore, we do not consider the temperature
effect on the nucleon-nucleon cross section in this work.

A conventional phase-space coalescence model \cite{Kru85} is used
to construct the clusters, in which the nucleons with relative
momenta smaller than $P_0$ and relative distances smaller than
$R_0$ are considered to belong to one cluster. In this work, $P_0$
and $R_0$ are chosen to be $0.3\,{\rm GeV}/c$ and $3.5$ fm. These
are the same values as in our previous works
\cite{{Li:2005zz},{Li:2005kq},Li:2005gf}. The freeze-out time is
taken to be $150\ {\rm fm}/c$.

\begin{figure}
\includegraphics[angle=0,width=0.6\textwidth]{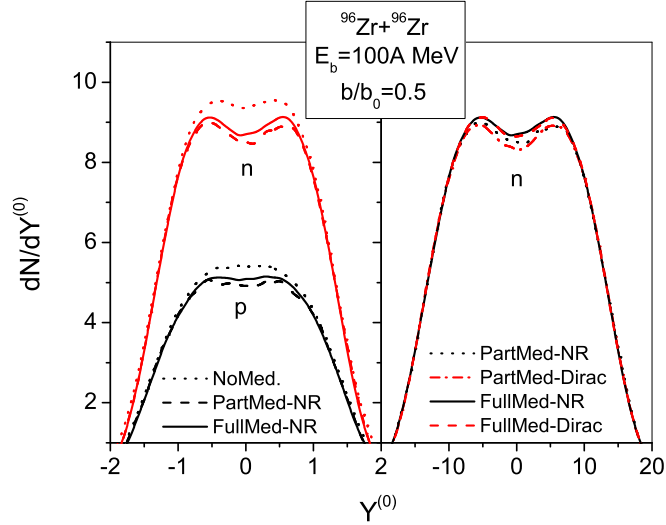}
\caption{Rapidity distributions of unbound neutrons and protons
for Zr+Zr reactions at $E_b=100A$ MeV and $b/b_0=0.5$. The left
plot compares the results of three cases ("NoMed", "PartMed-NR",
and "FullMed-NR") for protons and neutrons. The right plot shows
the NR and Dirac splitting effects on neutrons for the constraints
"PartMed" and "FullMed" (see context).} \label{figr1}
\end{figure}

Fig.\ \ref{figr1} shows the rapidity ($Y^{(0)}=y_{c.m.}/y_{beam}$)
distribution of unbound neutrons and protons for
$^{96}$Zr+$^{96}$Zr reactions (initial $\alpha\simeq 0.167$) at
beam energy $E_b=100A$ MeV and reduced impact parameter
$b/b_0=0.5$. In the left part, the calculations of the three cases
("NoMed", "PartMed-NR", and "FullMed-NR") are compared for unbound
protons and neutrons. It is seen that the yields of nucleons at
midrapidity are indeed influenced by the medium modification of
the NN elastic cross sections. This is understandable since these
nucleons are emitted mainly from the high density region where the
NN cross sections are strongly reduced. Due to the reduction of
the cross section, less nucleons are emitted for the cases
"PartMed" and "FullMed". Slightly more nucleons are emitted for
the case of "FullMed-NR" compared to the case of "PartMed-NR"
because the medium correction factor $F_{u}^p$ increases with
momentum as shown in Eq. (\ref{fdpup}). Due to the neutron-rich
environment, the rapidity distribution of neutrons is strongly
influenced by the medium modification of the two-body cross
section than that of protons. Therefore, in Fig.\ \ref{figr1}
(right), the different splitting effects of cross sections (Dirac
and NR) are only shown for unbound neutrons. It is found that in
general the splitting effect on the rapidity distribution of
neutrons is small and can only be detected for the case "PartMed",
while it is almost negligible for the case "FullMed". This is
because the splitting effect decreases strongly with increasing
momentum (see Eq. (\ref{fdpup})). For "PartMed" cases and at
midrapidities, the unbound nucleons of Dirac splitting are a
little smaller than those of NR due to a further reduction of the
$\sigma^*_{nn}$ for the Dirac case compared to NR case in the
nuclear medium.

\begin{figure}
\includegraphics[angle=0,width=0.6\textwidth]{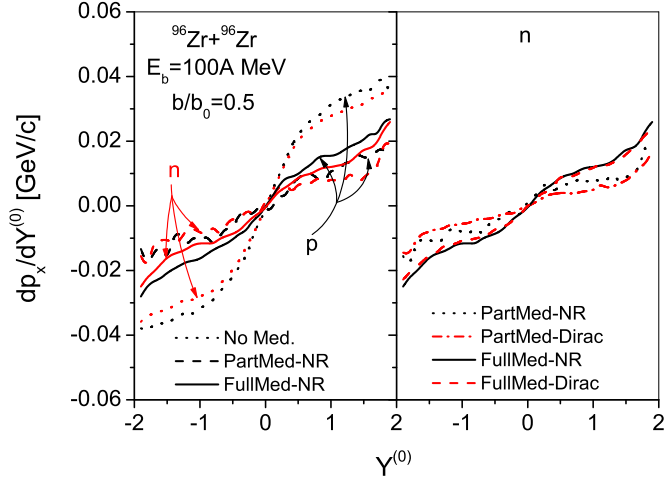}
\caption{Transverse flow distribution of unbound neutrons and
protons as a function of rapidity for Zr+Zr reactions at
$E_b=100A$ MeV and $b/b_0=0.5$. In the left plot, the results of
three cases ("NoMed", "PartMed-NR", and "FullMed-NR") are compared
for protons and neutrons. The right plot shows the NR and Dirac
splitting effects on neutron flow for "PartMed" and "FullMed".}
\label{figr2}
\end{figure}

Transverse flow was proposed as a sensitive probe of the medium NN
elastic cross section \cite{Gyu82etc}. Here we show the in-plane transverse flow of nucleons in
Fig.\ \ref{figr2} as a function of rapidity. The same trend as in
Fig.\ \ref{figr1} with respect to the medium correction of cross
sections is found. In Fig.\ \ref{figr2} (left) the case "NoMed"
produces the largest positive flow for unbound nucleons, while the
case "PartMed" shows the smallest flow and "FullMed" lies
in-between. Due to the Coulomb potential on protons, the positive
transverse flow of protons is always higher than that of neutrons.
It is known that above the balance energy, the repulsively
nucleon-nucleon scattering effect gains increasing importance,
which leads to a positive flow parameter. Similarly, Fig.\
\ref{figr2} (right) indicates that the unbound neutron flows
calculated with the different mass splitting effect of NR and
Dirac, leads only in the "PartMed" case to a difference of neutron
flows while there is no difference for the "FullMed" case. This
strong effect of the medium modified cross sections on the
dynamics of HICs but the small difference corresponding to
different mass splitting of NR and Dirac mass is easy to
understand because the splitting effect due to the isovector part
is rather small compared to the isoscalar density effects. This
finding is consistent with Ref.\ \cite{Li:2005jy} with respect to
the splitting effect in NN cross section shown in Figs.\
\ref{figr1} and \ref{figr2}. This feature of transverse flow is
very important to experimentally measure the medium modification
of cross sections. However, on the one hand, it is known that the
uncertainties of iso-scalar part of the mean-field potentials has
also strong effect on the transverse flow, while on the other
hand, it will be difficult to discover the probable splitting of
neutron-neutron and proton-proton effective cross sections in
isospin-asymmetric nuclear medium. Thus, one needs further
observables for testing in-medium cross sections to obtain
unambiguous conclusions.

\begin{figure}
\includegraphics[angle=0,width=0.6\textwidth]{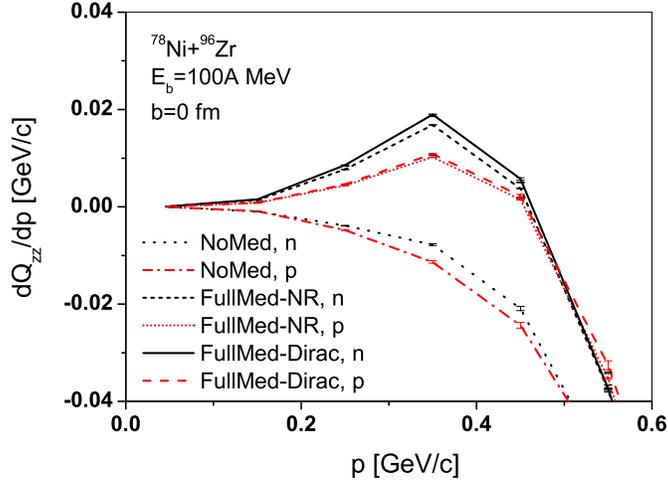}
\caption{Momentum distribution of $Q_{zz}$ of neutrons and protons
separately. Two cases "NoMed" and "FullMed" (with NR or Dirac
splitting effect) are chosen for $^{78}$Ni+$^{96}$Zr reactions at
$b=0$ fm and $E_b=100A$ MeV.} \label{figr3}
\end{figure}

The momentum quadrupole $Q_{zz}$ ($=\langle 2p_z^2-p_x^2-p_y^2
\rangle$), which is usually taken to measure the stopping power,
has also been extensively studied as a good messenger of the
medium modifications of elastic NN cross section. Its major
advantage is the weak dependence on the uncertainties of the
symmetry energy (for example, see Refs.
\cite{Liu:2001uc,Li:2002ag,Chen:2003wp}). Fig.\,\ref{figr3} shows
the momentum distributions of the average $Q_{zz}$ of neutrons and
protons calculated for two cases "NoMed" and "FullMed"
(considering NR and Dirac splitting effects) for
$^{78}$Ni+$^{96}$Zr reactions (initial $\alpha\simeq0.218$) at
$b=0$ fm and $E_b=100A$ MeV. At first glance, we see that the
result for the case "NoMed" is negative in the whole momentum
region, while the results for the case "FullMed" with different
splitting effects are positive at low momenta $p<0.5\,{\rm GeV}/c$
and return to be negative at larger momenta. From the definition
of $Q_{zz}$ it is easy to see that positive $Q_{zz}$ values mean
incomplete stopping or nuclear transparency while negative
$Q_{zz}$ values mean transverse expansion or collectivity. Hence,
the result with free cross sections shows strong collectivity in
the whole momentum region (which is in line with the transverse
in-plane flow, see Fig.\ \ref{figr2}), while the results with
medium modifications (i.e., with reduced cross sections) show
larger transparency. The conversion of $Q_{zz}$ (after integrating
the whole momentum space) from negative to positive was also seen
\cite{Chen:2003wp} and should be easily measurable by experiments.
From Fig.\ \ref{figr3} we also find that the $Q_{zz}$ of neutrons
at moderate momenta ($\sim 0.15 - 0.45\,{\rm GeV}/c$) is always
larger than that of protons, while this difference disappears at
larger momenta. In our previous work \cite{Li:2005zz} we found
that the average collision number of neutrons is always smaller
than that of protons for neutron-rich intermediate-energy HICs,
which implies that the neutrons experience larger transparency
than protons. Furthermore, the $Q_{zz}$ result of neutrons with
Dirac splitting effect shows a larger transparency than that with
the NR one, especially at moderate momenta $\sim 0.35{\rm GeV}/c$
due to a smaller $\sigma_{nn}^*$ in a neutron-rich system. In the
analysis of Ref.\,\cite{Li:2005ib} it was shown that $Q_{zz}$ is
only sensitive to the magnitude of the cross section but not the
isospin dependence of the NN cross sections. In this work we
further find that the difference between two different medium
corrections due to different mass splittings is rather small
because the magnitude difference of cross sections caused by
different mass splitting of NR and Dirac we studied is rather
small compared to the cross section itself.

\begin{figure}
\includegraphics[angle=0,width=0.6\textwidth]{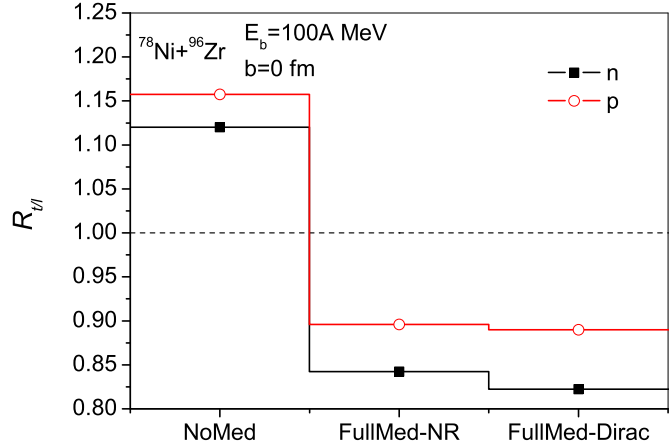}
\caption{The $R_{t/l}$ values of unbound neutrons and protons with
different medium modifications on cross sections ("NoMed",
"FullMed-NR", and "FullMed-Dirac"). The reaction
$^{78}$Ni+$^{96}$Zr at $b=0$ fm and $E_b=100A$ MeV is chosen (see
context).} \label{figr4}
\end{figure}

In order to measure the degree of the stopping, the ratio of
variances of the transverse (x-axis) to that of the longitudinal
(z-axis) rapidity distribution, called vartl, was recently
proposed and measured \cite{Reisdorf:2004wg}. Alternatively and
similarly, here we define
$R_{t/l}=\Gamma_{dN/dY_x}/\Gamma_{dN/dY_z}$ where $\Gamma$ means
the halfwidth of the rapidity distribution. In a thermal
nonequilibrated system, the $R_{t/l}$ value will depart from unit:
super-stopping leads to $R_{t/l} > 1$ while large transparency
leads to $R_{t/l} < 1$. Fig.\ \ref{figr4} shows the calculation
results for $R_{t/l}$ values of neutrons and protons with
different medium modifications on cross sections. One clearly
observes a large effect (more than $20\%$) of the medium
modification of the NN cross section on $R_{t/l}$ (from the case
"NoMed" to "FullMed"). But the difference between the results
calculated with "FullMed-NR" and "FullMed-Dirac" is still very
small. Here the effect is only about $2\%$.

\begin{figure}
\includegraphics[angle=0,width=0.6\textwidth]{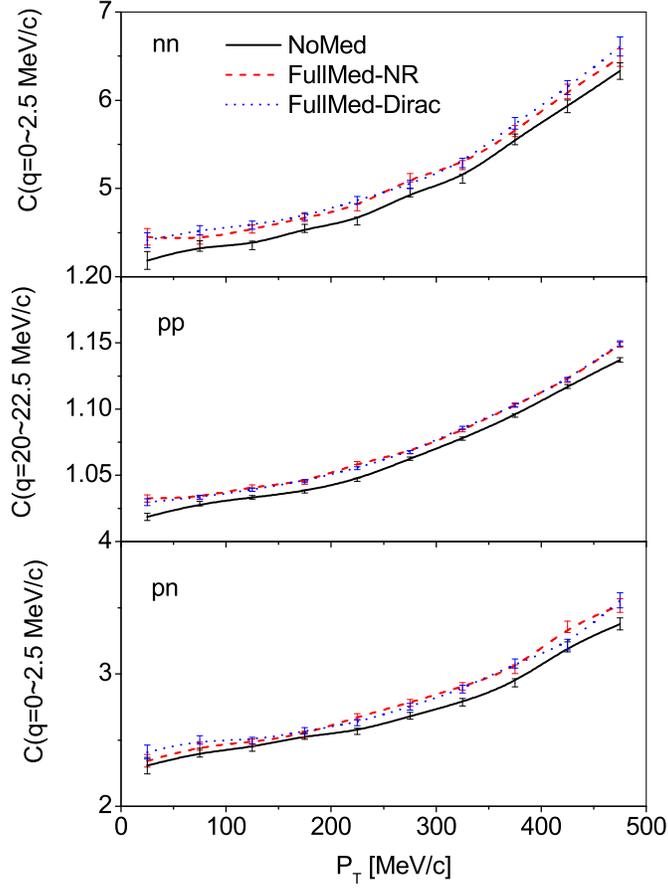}
\caption{The n-n (top), p-p (middle) and p-n (bottom) correlation
functions as a function of total transverse momentum $P_T$ and for
fixed relative momentum $q$ without and with medium modifications
(NR and Dirac splittings) of the nucleon-nucleon elastic cross
sections. $^{78}$Ni+$^{96}$Zr reactions at $b=0$ fm and $E_b=100A$
MeV are chosen.} \label{figr5}
\end{figure}

Recently, Chen et al \cite{Chen:2003wp} found that the two-nucleon
correlation functions in neutron-rich intermediate energy HICs are
sensitive to the density dependence of the nuclear symmetry
energy. Soon afterwards the isospin effects were investigated
experimentally in the $E_b=61A$ MeV $^{36}$Ar+$^{112,124}$Sn
reactions by Ghetti et al \cite{Ghetti:2003pv} and it seems that
the two-particle correlation functions is indeed a useful probe
for the isospin dependence of the nuclear EoS. It was also pointed
out that the two-body correlation is unsensitive to both the
isoscalar part of the EoS and in-medium cross sections \cite
{Chen:2003wp}. In this work we elaborate on the sensitivity of the
two-particle correlation function to the in-medium NN cross
section. To calculate the unbound NN correlation functions, we
adopt the Koonin-Pratt method. The program Correlation After
Burner (CRAB) (version 3.0) is used, which is based on the formula
\cite{{Pratt:1994uf},{Pratt:1986ev},{Pratt:1984su},Koonin:1977fh}:
\begin{equation}
C({\bf P},{\bf q}) = \frac {\int d^4x_1 d^4x_2 g(x_1,{\bf P}/2)
g(x_2,{\bf P}/2) |\phi({\bf q}, {\bf r})|^2} {{\int d^4x_1
g(x_1,{\bf P}/2)} {\int d^4x_2 g(x_2,{\bf P}/2)}}. \label{cpq}
\end{equation}
Here $g(x,{\bf P}/2)$ is the probability for emitting a particle
with momentum ${\bf P}/2$ from the space-time point $x = ({\bf r},
t)$. $\phi({\bf q}, {\bf r})$ is the relative two-particle wave
function with ${\bf r}$ being their relative position. ${\bf
P}={\bf p}_1+{\bf p}_2$ and ${\bf q}=({\bf p}_1-{\bf p}_2)/2$ are
the total and relative momenta of the particle pair, respectively.
According to previous studies (see for examples,
\cite{Bauer:1993wq,Ghetti:2000ab,Chen:2003wp}), the effect of
nuclear medium on neutron-neutron, neutron-proton correlation
functions is more pronounced at smaller relative momentum ($q$)
and larger total momentum ($P$) than that with larger $q$ or
smaller $P$ and that on proton-proton correlation function is more
pronounced at $q\sim 20 {\rm MeV}/c$ and with larger $P$. In
Fig.\, \ref{figr5} we show the transverse momentum $P_T$
($=\sqrt{(p_{1x}+p_{2x})^2+(p_{1y}+p_{2y})^2}$) dependence of the
correlation functions ($C(q,P_T)$) for neutron-neutron (top),
proton-neutron (bottom) pairs within the bin of relative momentum
$q=0\sim 2.5 {\rm MeV}/c$ and for proton-proton (middle) pairs
within the bin $q=20\sim 22.5\, {\rm MeV}/c$. For better
precision, $10^9$ neutron-neutron and proton-neutron pairs and
$10^8$ proton-proton pairs are analyzed for $P_T<500\, {\rm
MeV}/c$ and $q < 50\, {\rm MeV}/c$. From Fig.\ \ref{figr5} one
sees that with the increase of $P_T$, the two-nucleons exhibit
enhanced correlation obviously due to the short average spatial
separation at the emission time. In the transverse momentum region
studied here, the results for the case "NoMed" are always lower
than for the other cases. The observed increase of the correlation
function with the reduction of the NN elastic cross sections was
also seen in \cite{Bauer:1993wq}. A small effect of the medium
correction of two-body cross sections on the two-particle
correlation is also seen, while the difference between the results
with NR and Dirac splitting is negligible. Thus the two-particle
correlation function is an ideal observable for probing the
density dependence of the nuclear symmetry energy.

In summary, we have investigated several observables concerning
unbound nucleons which are  to some extent sensitive to the medium
modifications of nucleon-nucleon elastic cross sections in
neutron-rich intermediate energy heavy ion collisions. The
splitting effect of neutron and proton effective masses on cross
sections has been discussed. Although many suggested observables,
such as the well-known rapidity and the momentum distributions of
the yields of nucleons, are sensitive to the medium modifications
of cross sections, they are also subject to the uncertainties in
the mean field part making the conclusions somehow ambiguous. The
transverse flow as a function of rapidity, and especially the
$Q_{zz}$ as a function of momentum and the ratio of halfwidths of
the transverse to that of longitudinal rapidity distribution
$R_{t/l}$ are found to be the highly sensitive probes of the medium
modifications of the cross sections. The transverse momentum
distributions of correlation functions of two-nucleons is
unsensitive to the cross sections. The difference between the
in-medium cross section modified by NR and Dirac effective
neutron- and proton-  mass splitting on these observables was
shown to be very small, suggesting to find more sensitive
observables to explore these effect.

\section*{Acknowledgments}
We would like to thank Scott Pratt for the use of the CRAB
program and acknowledge support by the Frankfurt Center for Scientific Computing (CSC). Q. Li thanks the Alexander von Humboldt-Stiftung for a
fellowship. This work is partly supported by the National Natural
Science Foundation of China under Grant No.\ 10235030 and the
Major State Basic Research Development Program of China under
Contract No. G20000774, as well as by GSI, BMBF, DFG, and
Volkswagenstiftung.


\begin{thebibliography}{99}

\bibitem{Han94} Yinlu Han, G.J. Mao, Z. Li, and Y. Zhuo,   Phys.\ Rev.\ C {\bf 50}, 961 (1994)\\

\bibitem{Bass98}S. A. Bass {\it et al.}, Prog. Part. Nucl. Phys. {\bf 41}, 255 (1998)\\

\bibitem{BaoAnBook01} {\it Isospin Physics in Heavy Ion Collisions at Intermediate
Energies}, edited by Bao-An Li and W. Udo Schroeder, NOVA Science Publishers, Inc., New York, 2001\\

\bibitem{baranRP} V. Baran, M. Colonna, V. Greco, M. DiToro, Phys. Rep. {\bf 410}, 235 (2005)\\

\bibitem{Li:1997px}
  B.~A.~Li, C.~M.~Ko and W.~Bauer,
  Int.\ J.\ Mod.\ Phys.\ E {\bf 7}, 147 (1998)

\bibitem{Li:2000sh}
  Q.~Li, Z.~Li and G.~Mao,
  Phys.\ Rev.\ C {\bf 62}, 014606 (2000)

\bibitem{Li:2003vd}
  Q.~Li, Z.~Li and E.~Zhao,
  Phys.\ Rev.\ C {\bf 69}, 017601 (2004)

\bibitem{Liu:2001iz}
  B.~Liu, V.~Greco, V.~Baran, M.~Colonna and M.~Di Toro,
  Phys.\ Rev.\ C {\bf 65}, 045201 (2002)

\bibitem{DiToro:2005ac}
  M.~Di Toro, M.~Colonna and J.~Rizzo,
  arXiv:nucl-th/0505013.

\bibitem{vanDalen:2005ns}
  E.~N.~E.~van Dalen, C.~Fuchs and A.~Faessler,
  Phys.\ Rev.\ Lett.\  {\bf 95}, 022302 (2005)

\bibitem{Sammarruca:2005ch}
  F.~Sammarruca,
  arXiv:nucl-th/0506081.

\bibitem{Sammarruca:2005tk}
  F.~Sammarruca and P.~Krastev,
  arXiv:nucl-th/0509011.

\bibitem{Zuo:2001bd}
  W.~Zuo, I.~Bombaci and U.~Lombardo,
  Phys.\ Rev.\ C {\bf 60}, 024605 (1999)

\bibitem{Zuo:2005hw}
  W.~Zuo, L.~G.~Cao, B.~A.~Li, U.~Lombardo and C.~W.~Shen,
  Phys.\ Rev.\ C {\bf 72}, 014005 (2005)

\bibitem{Li:2005ib}
  B.~A.~Li, P.~Danielewicz and W.~G.~Lynch,
  Phys.\ Rev.\ C {\bf 71}, 054603 (2005)

\bibitem{Li:2005jy}
  B.~A.~Li and L.~W.~Chen,
  arXiv:nucl-th/0508024.

\bibitem{Li:2005zz}
  Q.~Li, Z.~Li, S.~Soff, R.~K.~Gupta, M.~Bleicher and H.~St\"ocker,
  J.\ Phys.\ G: Nucl.\ Part.\ Phys.\ {\bf 31}, 1359 (2005)


\bibitem{Li:2005kq}
  Q.~Li, Z.~Li, S.~Soff, M.~Bleicher and H.~St\"ocker,
  Phys.\ Rev.\ C {\bf 72}, 034613 (2005)

\bibitem{Li:2005gf}
  Q.~Li, Z.~Li, S.~Soff, M.~Bleicher and H.~St\"ocker,
  J.\ Phys.\ G: Nucl.\ Part.\ Phys.\ {\bf 32}, 151 (2006)

\bibitem{Alm:1994db}
  T.~Alm, G.~Ropke and M.~Schmidt,
  Phys.\ Rev.\ C {\bf 50}, 31 (1994)

\bibitem{Kru85}H. Kruse, B.V. Jacak, J.J. Molitoris, G.D. Westfall, H. St\"ocker, Phys. Rev. C {\bf 31}, 1770 (1985)\\

\bibitem{Gyu82etc}M. Gyulassy, K.A. Fraenkel, and H. St\"ocker, Phys. Lett. {\bf 110B}, 185 (1982);
               G. Peilert, H. St\"ocker, and W. Greiner, Phys. Rev. C {\bf 39}, 1402 (1989);
	       for reviews, see:
	       H. St\"ocker and W. Greiner, Phys. Rep. {\bf 137}, 277 (1986);
	       S. Bass, M. Gyulassy, and H. St\"ocker, J. Phys. G: Nucl.\ Part.\ Phys.\ {\bf 25}, R21 (1999)\\

\bibitem{Liu:2001uc}
  J.~Y.~Liu, W.~J.~Guo, S.~J.~Wang, W.~Zuo, Q.~Zhao and Y.~F.~Yang,
  Phys.\ Rev.\ Lett.\  {\bf 86}, 975 (2001)

\bibitem{Li:2002ag}
  Q.~Li and Z.~Li,
  Chin.\ Phys.\ Lett.\  {\bf 19}, 321 (2002)


\bibitem{Chen:2003wp}
  L.~W.~Chen, V.~Greco, C.~M.~Ko and B.~A.~Li,
  Phys.\ Rev.\ C {\bf 68}, 014605 (2003)

\bibitem{Reisdorf:2004wg}
  W.~Reisdorf {\it et al.}  [FOPI Collaboration],
  Phys.\ Rev.\ Lett.\  {\bf 92}, 232301 (2004)

\bibitem{Ghetti:2003pv}
  R.~Ghetti {\it et al.},
  Phys.\ Rev.\ C {\bf 69}, 031605 (2004)



\bibitem{Koonin:1977fh}
  S.~E.~Koonin,
  Phys.\ Lett.\ B {\bf 70} (1977) 43.

\bibitem{Pratt:1984su}
  S.~Pratt,
  Phys.\ Rev.\ Lett.\  {\bf 53} (1984) 1219.

\bibitem{Pratt:1986ev}
  S.~Pratt,
  Phys.\ Rev.\ D {\bf 33} (1986) 72.

\bibitem{Pratt:1994uf}
  S.~Pratt {\it et al.},
  Nucl.\ Phys.\ A {\bf 566} (1994) 103C.


\bibitem{Ghetti:2000ab}
  R.~Ghetti {\it et al.},
Nucl.\ Phys.\ A {\bf 674} (2000) 277.

\bibitem{Bauer:1993wq}
  W.~Bauer, C.~K.~Gelbke and S.~Pratt,
  Ann.\ Rev.\ Nucl.\ Part.\ Sci.\  {\bf 42} 77 (1992).




\end{thebibliography}
\end{document}